\documentclass[10pt,conference]{IEEEtran}
\IEEEoverridecommandlockouts

\usepackage{cite}
\usepackage{amsmath,amssymb,amsfonts}
\usepackage{algorithmic}
\usepackage{graphicx}
\usepackage{textcomp}
\usepackage{xcolor}
\usepackage{graphicx}
\usepackage{multirow}
\usepackage{booktabs}
\usepackage{array}
\usepackage{amssymb}  
\usepackage{mdframed} 
\usepackage{caption}
\usepackage{hyperref} 
\usepackage{tikz}
\usepackage{makecell}
\usepackage{xcolor}
\usepackage{float}
\usepackage{afterpage}
\usepackage[most]{tcolorbox}

\tikzstyle{startstop} = [rectangle, rounded corners, minimum width=2cm, minimum height=1cm, text centered, draw=black, font=\footnotesize]
\tikzstyle{process} = [rectangle, minimum width=2cm, minimum height=1cm, text centered, draw=black, font=\footnotesize]
\tikzstyle{decision} = [diamond, minimum width=2cm, minimum height=1cm, text centered, draw=black, font=\footnotesize]
\tikzstyle{arrow} = [thick,->,>=stealth]
\usetikzlibrary{shapes.geometric, arrows}

\def\BibTeX{{\rm B\kern-.05em{\sc i\kern-.025em b}\kern-.08em
    3T\kern-.1667em\lower.7ex\hbox{E}\kern-.125emX}}
    
\begin{document}

\title{Reducing Alert Fatigue via AI-Assisted Negotiation: A Case for \texttt{Dependabot}}

\author{%
Raula Gaikovina Kula, \\
\textit{ Osaka University, Japan}\\
raula-k@ist.osaka-u.ac.jp
}

\maketitle

\begin{abstract}
The increasing complexity of software dependencies has led to the emergence of automated dependency management tools, such as Dependabot.
However, these tools often overwhelm developers with a high volume of alerts and notifications, leading to alert fatigue.
This paper presents a position on using Artificial Intelligence (AI) agents as dependency negotiators to reduce alert fatigue.
We then examine specific use cases where AI agents can facilitate dependency negotiations, such as when working with external dependencies or managing complex, multi-component systems.
Our findings highlight the need for more research on the design and evaluation of AI-driven dependency mediation mechanisms. With a focus on ensuring transparency, explainability, and human trustworthiness in these GitHub software projects, our goal is to reduce alert fatigue to an extent that maintainers no longer feel overwhelmed and welcome pull requests just like any other contribution into their projects.
\end{abstract}


\section{Introduction}

It is not too far-fetched to say that GitHub's Dependabot\footnote{\url{https://docs.github.com/en/code-security/dependabot/dependabot-security-updates/about-dependabot-security-updates}} is probably the most widespread and well-known bot in the GitHub universe.
Such notoriety can be attributed to the attention needed to keep open-source projects free of security vulnerabilities.
As a dependency management bot, Dependabot became a key step in using automation to address the issues of updating dependencies.

Related work all show both the drawbacks and benefits of using bots, especially in the context of keeping third-party libraries updated and free of known security vulnerabilities.
For instance, some studies demonstrate that Dependabots are well-received by open-source projects\cite{alfadel2021use, he2023automating, mohayeji2023investigating, rebatchi2024dependabot} but sometimes with mixed feelings.
These studies show that Dependabot may only be considered for relatively easier fixes where very little manual effort is required.
Other work \cite{jernestaal2024evolving} suggests that Dependabots are often not resolved, as they become superseded.
Cogo \cite{cogo2022understanding} argues that bots require much configuration to be perfect.
Finally, He et al. \cite{he2023automating} propose that the ideal dependency management tooling requires four dimensions of configurability, autonomy, transparency, and self-adaptability

Apart from these issues, these related literature are in agreement that Dependabot has the unavoidable side-effect of generating noise in their updates.
For instance, in 2022, Dependabot automatically generated over 75 million pull requests to address millions of specific vulnerabilities\footnote{\url{https://www.forbes.com/sites/rogerdooley/2024/01/04/ai-in-negotiations-a-game-changer-for-dealmakers/}} .
Furthermore, related studies have noted that developers have dealt with fatigue alert issues.
To reduce such noise, Dependabot is getting smarter—and quieter—by reducing bot-based noise from repositories based on user interaction with Dependabot.
According to GitHub, the Dependabot has been improved to address these issues by being smarter with better options like automatically ceasing the creation of pull requests in inactive repositories\footnote{\url{https://github.blog/security/supply-chain-security/a-smarter-quieter-dependabot/}} .
Furthermore, maintainers of Dependabot have created a new "allow auto-dismissal" function that safely reduces the volume of false positive alerts, which can overwhelm developers and distract from legitimate vulnerabilities.


GitHub also noted that Dependabot will now use a set of rules to become smarter, thus reduce false positives.
\begin{quote}
   \textit{ “Rather than over-index on one criterion like reachability or dependency scope, we’ve designed an alert rules engine that uses a rich set of complex, contextual alert metadata. This way, Dependabot can relieve alert fatigue while remaining vigilant about alerts that might put your software at risk.”}
\end{quote}
Futhermore a blog post by GitHub stated that, \textit{
“By detecting and auto-dismissing false positives, today’s release will reduce the volume of npm alerts by approximately 15\%, and marks the beginning of a series of ships that improve the relevance of alerts and relieve alert fatigue.”}

This position paper proposes that developers can leverage generative AI so that an AI-assistant can be leveraged to become the negotiator between the Dependabot.
The arguement is that AI technology advancements can be leveraged as a broker for dependency updates. 
The key idea is that the AI agent will act on the behalf of the maintainer, and based on perspectives of the maintainer decide when to let an alert from dependabot pass through.
We envision the AI to sit between the dependabot and the maintainer, ensuring that an alert is only triggered if the AI deems that the security update is applicable to the software project.
Currently, the only requirement for a dependabot to alert a project is whether or not a project uses a dependency, yet we believe that there should be more nuance behind releasing alerts.

\section{The Role Of AI In Negotiating \\ When to Allow an Alert}
The idea of leveraging AI negotiation support systems \cite{doi:10.1177/20555636241269270} is based on the concept of assisting human negotiators by analyzing vast amounts of data, predicting potential outcomes, and offering valuable insights. 
Applying to the Dependabot case, the AI will be capable of providing the following:

\begin{enumerate}
    \item \textbf{Risk Assessment and Scenario Planning:}
    AI negotiation should excel at risk assessment and scenario planning. When using AI agents, they can attempt to apply the fix and assess its potential fit. Furthermore, the agent can scan other solutions within the ecosystem, anticipate challenges, and devise contingency plans.
    \begin{enumerate}
        \item Is the mitigation plan an upgrade, downgrade, drop or migration to alternative?
        \item How much source code changes is required to integrate the migration?
        \item Has the maintainer show a history of making these changes?
    \end{enumerate}
    
    \item \textbf{Data from ecosystem:}
    Inspired by the concept of leveraging information from the crowd\cite{rombaut2024leveraging}, one of the primary advantages of the AI agent will be its ability to process and analyze vast volumes of historical and real-time data, especially if the target project is part of an ecosystem of software projects that most projects belong to, such as NPM, PyPI, and others.
    The AI can mine this data to look for trends, analyze previous negotiation outcomes, and decisions made by other software projects that are in similar situations. For example, the AI might detect that most similar projects are updating their libraries, which could help inform a decision to allow an alert to pass through to the maintainer.
    
    \begin{enumerate}
        \item How much of the ecosystem reacts to the alert with a change?
        \item How quick was the security advisory life-cycle?
    \end{enumerate}
    
    \item \textbf{Real-time Feedback:}
    The key to the success of recent generative AI technologies like Large Language Models has been their ability to quickly respond in close to real-time.
    As time is of the essence, decisions need to be made promptly. AI-powered negotiation support systems can process real-time data, which enables them to provide timely and up-to-date information at the fingertips of the maintainer. A notice from Dependabot might trigger a real-time data analysis requirement, necessitating prompt processing to deliver relevant insights.
    \begin{enumerate}
        \item How quick can the AI agent reason and provide a decision to alert the maintainer?
    \end{enumerate}
    
\end{enumerate}

\section{Integrating Perspectives for a Solution}

As part of the negotiation, the AI framework will require both datasets and perspectives of the security threat, the response of the community, and the workload of the maintainer in its decision.
We outline two key perspectives below:

\begin{itemize}
    \item \textbf{External Users, Security, Similar projects:}
Data analytics platforms integrated with AI algorithms analyze extensive historical and real-time data relevant to the negotiation.
They provide negotiators with valuable insights from GitHub, as well as other online resources such as StackOverflow, Reddit, and Hacknews.
The AI agent should also be able to scan security databases, including the GitHub Advisory, Synk, and official NVD and CVE websites. Additionally, the solution should be able to learn from the history of similar Dependabot actions from other similar projects.

\item \textbf{Internal Maintainer Interactions:}
AI agents like ChatGPT demonstrate the success of using interactive conversations as the best means to interface with humans.
Therefore, we envision this AI agent to interface with the maintainer in an interactive conversation, understanding the maintainer's perspectives on what types of alerts would be appreciated and which would cause fatigue.
The AI agent should be able to gain the trust of the maintainer to alert with the best interest of the maintainer in mind.
\end{itemize}

\section{A Call for Solutions}

Addressing the issue of maintainer fatigue from dependency updates through bots could be a practical application of AI assistance in software development tasks. Technologies such as large language models (LLMs) and AI agents are both timely and have the potential to enhance dependency management by incorporating diverse perspectives. However, these innovations come with three key challenges for a solution.
One of the primary challenges will be establishing trust between the AI agent and the maintainer. It will also be crucial to determine the extent to which the AI agent can offer reasonable recommendations without generating false or misleading claims. Furthermore, it is important to ensure that maintainers can verify that the AI agent is not omitting critical security vulnerability fixes.
The second challenge is ensuring that datasets are kept up-to-date while maintaining a real-time response to any Dependabot requests.
The final challenge involves understanding potential biases within the system and ensuring that the maintainer’s key interests are always prioritized.

\bibliographystyle{ieeetr}
\bibliography{bibliography}

\end{document}